\documentclass[a4paper]{article}
\usepackage{footmisc}
\usepackage{ISCSLP2022}
\usepackage{amsfonts}
\usepackage{graphicx}
\usepackage{multirow} 
\usepackage{subfigure} 
\usepackage{hyperref}
\usepackage{enumitem}
\hypersetup{hidelinks,
	colorlinks=true,
	allcolors=black,
	pdfstartview=Fit,
	breaklinks=true}

\title{SAMOS: A Neural MOS Prediction Model Leveraging Semantic Representations and Acoustic Features}
\name{Yu-Fei Shi, Yang Ai$^*$, Ye-Xin Lu, Hui-Peng Du, Zhen-Hua Ling}
\address{National Engineering Research Center of Speech and Language Information Processing, \\University of Science and Technology of China, Hefei, P. R. China }
\email{zkddsr2023@mail.ustc.edu.cn, yangai@ustc.edu.cn, yxlu0102@mail.ustc.edu.cn, redmist@mail.ustc.edu.cn,  zhling@ustc.edu.cn}
\begin{document}
\maketitle

\begin{abstract}
Assessing the naturalness of speech using mean opinion score (MOS) prediction models has positive implications for the automatic evaluation of speech synthesis systems. 
Early MOS prediction models took the raw waveform or amplitude spectrum of speech as input, whereas more advanced methods employed self-supervised-learning (SSL) based models to extract semantic representations from speech for MOS prediction. 
These methods utilized limited aspects of speech information for MOS prediction, resulting in restricted prediction accuracy.
Therefore, in this paper, we propose SAMOS, a MOS prediction model that leverages both Semantic and Acoustic information of speech to be assessed. 
Specifically, the proposed SAMOS leverages a pretrained wav2vec2 to extract semantic representations and uses the feature extractor of a pretrained BiVocoder to extract acoustic features. 
These two types of features are then fed into the prediction network, which includes multi-task heads and an aggregation layer, to obtain the final MOS score.
Experimental results demonstrate that the proposed SAMOS outperforms current state-of-the-art MOS prediction models on the BVCC dataset and performs comparable performance on the BC2019 dataset, according to the results of system-level evaluation metrics.

\end{abstract}

\section{Introduction}

Text-to-speech (TTS) synthesis and voice conversion (VC) are focal points in the field of speech research, where the evaluation of the quality of speech synthesized by TTS and VC systems involves both objective and subjective assessments. 
Common objective evaluation metrics, such as mel-cepstral distance (MCD) \cite{kubichek1993mel} and signal-to-noise ratio (SNR) \cite{le2019sdr}, have limited correlation with human perception of speech quality. 
As a result, some objective measures or models related to human perception have been proposed \cite{rix2001perceptual,taal2010short,hines2015visqol,mittag2021nisqa}. %, such as perceptual evaluation of speech quality (PESQ) \cite{rix2001perceptual}, short-time objective intelligibility (STOI) \cite{taal2010short}, virtual speech quality objective listener (VISQOL) \cite{hines2015visqol}, and non-intrusive speech quality assessment NISQA \cite{mittag2021nisqa}. Intrusive methods, like PESQ, require a reference signal, making it impractical for evaluating synthesized speech. Non-intrusive methods like NISQA focus on predicting the quality of wideband speech transmission and are commonly used to evaluate codec quality. 
However, these objective methods typically require reference speech, making it impractical for evaluating synthesized speech signals. 
Therefore, for synthesis systems, subjective evaluation through listening tests is considered as the gold standard, with mean opinion score (MOS) testing being a common practice. 
In MOS testing, each listener is asked to rate the naturalness of a speech sample on a scale from 1 to 5. 
Due to the time-consuming and expensive nature of MOS testing, developing automated MOS prediction methods is necessary.

\graphicspath{ {plot/} }
\begin{figure*}[htbp]
    \setlength{\belowcaptionskip}{-0.4cm}
    \centering
    \includegraphics[width=17cm,height=5cm]{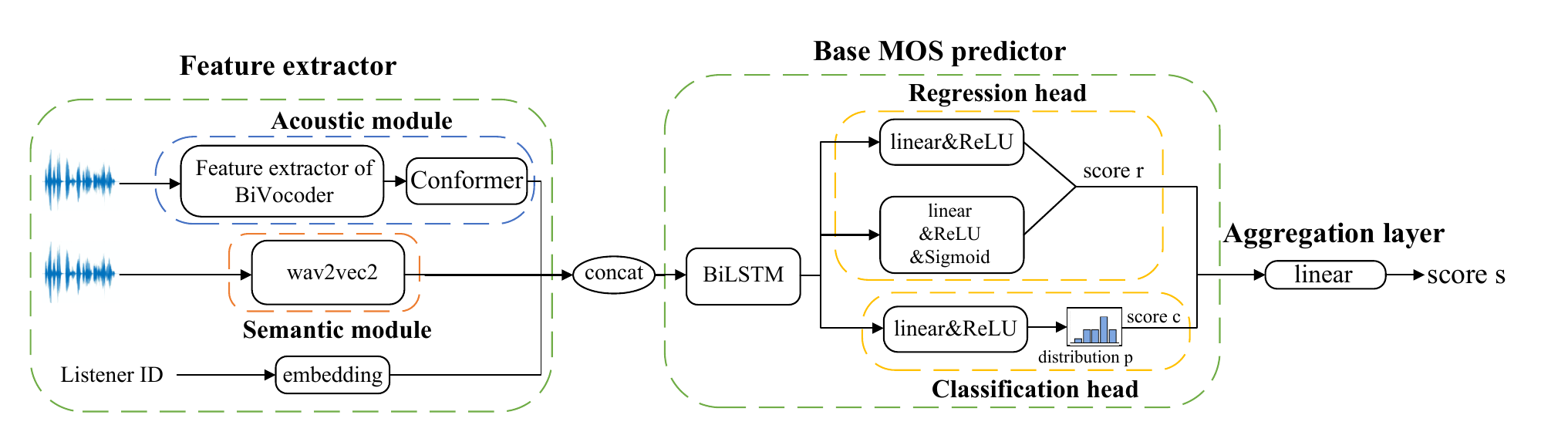}
    \caption{Overall structure of the proposed SAMOS model, where the ``concat" represents the feature concatenation operation, \textit{p} represents the probability scores of various classes outputted by the classification head, \textit{r} and \textit{c} represent the scores outputted by the regression head and classification head respectively, and \textit{s} represents the final score.}
    \label{fig:system_structure}
\end{figure*}

Early MOS prediction models adopted bidirectional long short-term memory (BiLSTM) based recurrent neural networks (RNNs) and convolutional neural networks (CNNs) for single-sentence score prediction \cite{patton2016automos,fu2018quality,lo2019mosnet} with amplitude-related features as input. 
%MOSNet \cite{lo2019mosnet} further improved performance %on the VC2016 and VC2018 datasets using a convolutional neural network-bidirectional LSTM (CNN-BiLSTM) structure and 
%by taking spectrum as input. 
These works all took the average of scores given by multiple listeners for a sentence as the target, however, different listeners actually provide different judgments for the same sentence. 
Realizing this, MBNet \cite{leng2021mbnet} and LDNet \cite{huang2022ldnet} considered listener information and added the scores from each listener as input, achieving some progress in prediction accuracy. 
Recently, with the rise of self-supervised-learning (SSL) based models trained on large-scale unlabeled data, fine-tuning SSL models and extracting SSL representationson on MOS datasets to leverage their high-level semantic information into MOS prediction models has become a widely used method, demonstrating impressive performance. 
Representative methods include SSL-MOS \cite{cooper2022generalization}, which added a linear layer on SSL model for fine-tuning on MOS datasets, and MOSA-Net \cite{zezario2022deep}, which used pretrained HuBERT \cite{hsu2021hubert} to extract SSL features and fed them into BiLSTM or CNN for MOS prediction. 
The VoiceMOS Challenge 2022 \cite{huang2022voicemos} was the first competition focused on MOS prediction tasks, where top-ranking systems mostly extended the SSL-MOS. 
Participants made several improvements into the baseline model, e.g., adding listener information embedding \cite{tian2022multi,tseng2022ddos,saeki2022utmos}, incorproating additional phoneme sequences extracted from ASR models \cite{saeki2022utmos}, and adopting multi-task learning \cite{tian2022multi} or ensemble learning \cite{saeki2022utmos,yang2022fusion,kunevsova2023ensemble}. 
After this challenge, some further improvement methods have gradually been proposed, such as integrating k-nearest neighbor (KNN) classification into SSL-based MOS prediction methods \cite{wang2023ramp} or introducing prosodic and linguistic features \cite{vioni2023investigating} as additional inputs to the model.
%\cite{saeki2022utmos,yang2022fusion,kunevsova2023ensemble}, and using ASR models to obtain evaluation scores or phoneme sequences, ensemble learning \cite{saeki2022utmos,yang2022fusion,kunevsova2023ensemble}.
% many improvements have been incorporated into the baseline include listener information embedding \cite{tian2022multi,tseng2022ddos,saeki2022utmos}, multi-task learning \cite{tian2022multi}, using ASR models to obtain evaluation scores or phoneme sequences, ensemble learning \cite{saeki2022utmos,yang2022fusion,kunevsova2023ensemble}. 
% Subsequent works introduced some new approaches, including k-nearest neighbor (KNN) classification \cite{wang2023ramp} and prosodic and linguistic features \cite{vioni2023investigating}.%Wang \textit{et al}. \cite{wang2023ramp} integrates k-nearest neighbor classification (KNN) and neural networks to enhance the model's generalization capability, Vioni \textit{et al}. \cite{vioni2023investigating} incorporates prosodic and linguistic features into the network, validating the effects on multiple baselines in new datasets.% Overall, existing state-of-the-art models mostly utilize semantic features of synthesized speech for MOS prediction.

However, most state-of-the-art models only utilized semantic features from SSL models or used a single-task framework to predict frame-level scores with the same weight, limiting the accuracy of MOS prediction.
Therefore, we introduce a novel neural MOS prediction model called SAMOS, which fully utilizes both semantic and acoustic information in speech. 
Semantic representations extracted from a pretrained wav2vec2 \cite{baevski2020wav2vec}, acoustic features extracted by a pretrained BiVocoder \cite{du2024bivocoder}, and listener-labeled information are jointly used as inputs to the model. 
The acoustic features contains compressed both amplitude and phase details, which provide more comprehensive information compared to amplitude-only related features commonly used in previous methods \cite{patton2016automos,fu2018quality,lo2019mosnet}. 
%magnitude spectrum and are compressed compared to raw waveform, reducing redundancy and irrelevant information. 
To further boost MOS prediction accuracy, the SAMOS also incorporates serveral techniques into its prediction network, such as multi-task learning heads \cite{tseng2022ddos}, weight branch \cite{shen2023sqat}, and aggregation layer.
%refinement layer \cite{tseng2022ddos}. 
Without using model ensemble strategy, the proposed SAMOS achieves state-of-the-art performance on the main track BVCC dataset \cite{cooper2021voices} of the VoiceMOS Challenge 2022 with a single neural network model. 
Additionally, the SAMOS also achieves comparable performance to other MOS prediction methods on the out-of-domain BC2019 dataset \cite{wu2019blizzard}.
%our model performs excellently on the out-of-domain dataset (BC2019) of the VoiceMOS Challenge 2022.

\vspace{-1.5mm}
\section{Proposed Method}
\vspace{-1.5mm}

The model structure of SAMOS is illustrated in Figure \ref{fig:system_structure}. 
First, the feature extractor produces three types of features, i.e., semantic representations derived from a wav2vec2-based semantic module \cite{baevski2020wav2vec}, acoustic features extracted by an acoustic module composed of the feature extractor of BiVocoder \cite{du2024bivocoder} and Conformers \cite{gulati2020conformer}, and listener-labeled information generated from the listener ID through a learnable embedding. 
We concatenate these three types of features along the feature dimension and feed them into the base MOS predictor. 
The base MOS predictor consists of a BiLSTM network, followed by parallel classification and regression heads, which respectively output the probability distribution $p$ of score classification and regression score $r$. 
We further take the expectation of the probability distribution as the classification score $c$, and feed both $c$ and $r$ into an aggregation layer to output the final MOS score $s$. 

\begin{figure}[t]
    \setlength{\belowcaptionskip}{-0.4cm}
    \centering
    \includegraphics[scale=0.4]{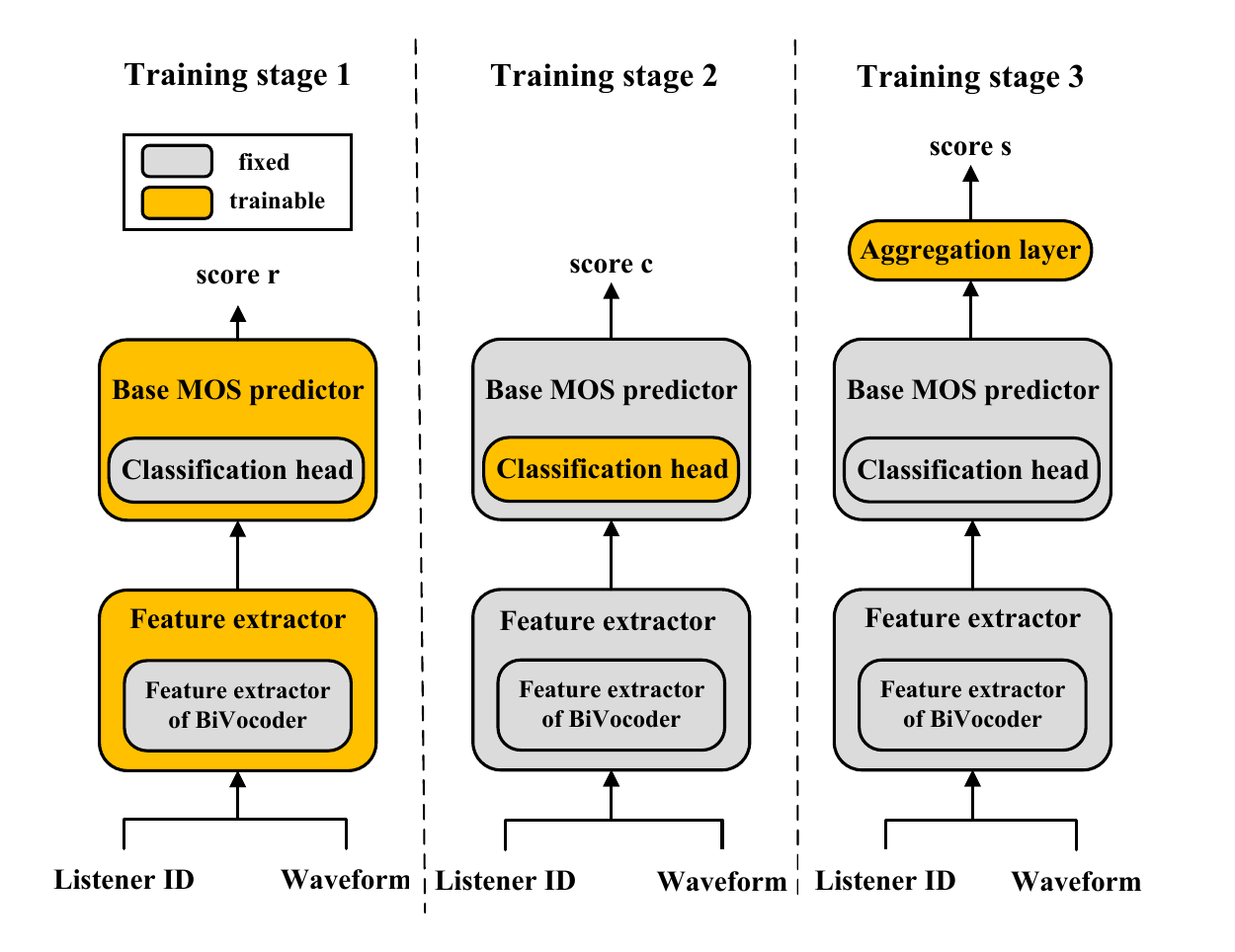}
    \caption{Training process of the proposed SAMOS.}
    \label{fig:procedure}
\end{figure}

We employ a stage-wise training mode for SAMOS as illusrated in Figure \ref{fig:procedure}. 
%The training process of SAMOS is illusrated in Figure \ref{fig:procedure}, which contains three stages: 
\begin{itemize}[nosep, leftmargin=*]
\item {}{\textbf{Training stage 1}}: We first freeze the parameters of pre-trained BiVocoder and classification head, and train other components in the feature extractor and the base MOS predictor to get score $r$ outputted by regression head.
\item {}{\textbf{Training stage 2}}: Then, we swap the frozen and trained modules (BiVocoder remains frozen) to obtain score $c$ outputted by classification head. 
\item {}{\textbf{Training stage 3}}: Finally, we freeze the whole feature extractor and base MOS predictor, and introduce the aggregation layer to train it separately, outputting the final score $s$.
\end{itemize}
%Subsequently, we will provide a detailed introduction to each module.
% $1)$ We freeze the parameters of pre-trained BiVocoder and classification head, and train feature extractor as well as base MOS predictor to get score $r$. $2)$ We freeze the parameters of other parts of feature extractor and base MOS predictor, and train the classification head to get score $c$. $3)$ We freeze the parameters of the entire model and add the refinement layer, training the refinement layer separately. We will detail each module in the following sections.

\subsection{Feature extractor}
\vspace{-1mm}
\subsubsection{Semantic module and listenr ID embedding}
\vspace{-1mm}
Following previous research settings \cite{cooper2022generalization}, we utilize a pre-trained SSL model (wav2vec2) to extract frame-level semantic representations from raw waveform. 
Different from SSL-MOS \cite{cooper2022generalization}, SAMOS processes the frame-level semantic representations through subsequent networks. 
To fully leverage the ratings given by each listener for a single sentence, inspired by \cite{huang2022ldnet}, we assign IDs to all raters and feed them into a learnable embedding to obtain listener features. It is worth noting that for a single sample $x_i$, there are individual scores $s_i^1,s_i^2,...,s_i^m$ given by $m$ listeners and the average score $\bar s$ across all listeners. Each individual score corresponds to the ID of the rater. We assign a virtual ‘‘mean-listener" ID to $\bar s$, so during training, a sample is trained $m+1$ times. During inference, since the rater of the sample is unknown, we use the mean-listener ID as input.
\graphicspath{ {plot/} }

\begin{table*}[t]

\setlength{\belowcaptionskip}{-0.4cm}
 \caption{The system-level performances of SAMOS along with five baselines on the test sets of BVCC (main track) and BC2019 (OOD track), respectively. The ``Y" after the model name indicates that the model adopts an ensemble strategy. The \textbf{bold} and \underline{underline} numbers indicate the optimal and sub-optimal results, respectively.}
 \label{table:main experiment}
 \centering
 \begin{tabular}{c| c c c c|c c c c} 
  \hline
  \multirow{2}{*}{}&\multicolumn{4}{c|}{BVCC}&\multicolumn{4}{c}{BC2019} \\
  \cline{2-9}
   & S\_MSE $\downarrow$ & S\_LCC $\uparrow$ & S\_SRCC $\uparrow$ & S\_KTAU $\uparrow$ & S\_MSE $\downarrow$ & S\_LCC $\uparrow$ & S\_SRCC $\uparrow$ & S\_KTAU $\uparrow$\\  
  \hline
   UTMOS (Y) & \textbf{0.090} & 0.939 & 0.936& \underline{0.794}& \textbf{0.030} & \textbf{0.988} & \textbf{0.979}& \textbf{0.908}\\
  T11 (Y) & 0.101 & \underline{0.941} & \underline{0.939}& \textbf{0.797}& \underline{0.048} & \underline{0.982} & 0.952& 0.852\\ 
   SSL-MOS & 0.113 & 0.928 & 0.923& 0.770& 0.093 & 0.971 & 0.975& 0.889\\ 
   UTMOS strong & 0.148 & 0.930 & 0.925& 0.774& 0.248 & 0.970 & 0.972& 0.879\\
   DDOS & \underline{0.091} & 0.940 & 0.938& 0.792& 0.070 & 0.960 & 0.969& 0.871\\
  
   SAMOS & 0.097 & \textbf{0.944} & \textbf{0.942}& \textbf{0.797}& 0.179 & 0.968 & \underline{0.976}& \underline{0.895}\\
  
  \hline
 \end{tabular}
\end{table*}

\vspace{-1mm}
\subsubsection{Acoustic module}
\vspace{-1mm}
The acoustic module consists of a BiVocoder and a Conformer. 
BiVocoder, which is our previous work \cite{du2024bivocoder}, is a newly proposed bidirectional neural vocoder with both feature extraction and waveform generation capabilities
%with its feature extractor using the ConvNeXt V2\cite{woo2023convnext} network as the backbone, which combines few parameters and strong modeling capabilities. 
Regarding the feature extraction module, speech amplitude and phase spectra are separately passed through a ConvNeXt v2 network \cite{woo2023convnext} and a downsampling layer, followed by concatenation operation and dimension reduction layer to obtain a compressed low-dimensional continuous feature containing both amplitude and phase information. 
% The amplitude spectrum and phase spectrum of speech processed through short-time Fourier transform (STFT) are separately passed through the ConvNeXt V2 block and then downsampled convolution, followed by concatenation and dimension reduction convolution to obtain a compressed representation containing both amplitude and phase information. 
This feature is then fed into the mirrored waveform generation module for speech reconstruction. 
In our work, we only use the feature extraction module of BiVocoder to extract acoustic features.
To further capture global information in acoustic features, we finally feed the output of the feature extraction module of BiVocoder into a Conformer to obtain the final acoustic features. 

\subsection{Base MOS predictor}
%The base MOS predictor receives the concatenated features for score prediction. 
The frame shift settings of wav2vec and BiVocoder are the same, so the number of frames for semantic and acoustic features is identical. 
Listener features are repeated along the time axis to match the same number of frames. 
These three features are concatenated along the dimension axis and injected into the base MOS predictor for MOS score prediction.
% Both the feature extractors of BiVocoder and wav2vec2 use a frame shift of $20ms$, making the semantic and acoustic features equal in the time dimension, denoted as $T$. 
% We replicate the listener features $T$ times in the time dimension, allowing the three features to be concatenated along the feature dimension. 
Inspired by \cite{tseng2022ddos}, we first pass the concatenated features through a BiLSTM and then generate two scores separately by the classification and regression heads. 

\vspace{-1.5mm}
\subsubsection{Regression head}
\label{subsec: Regression}
\vspace{-1.5mm}
Serveral works \cite{saeki2022utmos,yang2022fusion} consider MOS prediction as a regression task, where the model directly outputs the MOS score for each frame. 
When aggregating the MOS score for a sentence, directly calculating the average of frame-level scores is not reasonable because the speech quality is not uniform across all frames. 
%In previous works by advanced systems \cite{saeki2022utmos,yang2022fusion}, MOS prediction is treated as a regression problem, where the linear layer in the final layer of the model outputs scores for multiple frames. 
% When calculating the predicted score for an utterance, the scores for each frame are averaged. 
% Since the quality of different frames of speech varies, the weights corresponding to different frames should also be different. 
Therefore, we adopt a two-branch structure \cite{shen2023sqat}, where one branch outputs scores for each frame and the other branch outputs corresponding weights for each frame. 
The final sentence-level regression score $r$ is obtained by performing a weighted average over scores of all frames.
%The final weighted frame-level scores are obtained by multiplying the corresponding parts of the two branches. 
We use the contrastive loss \cite{saeki2022utmos} and clipped loss \cite{leng2021mbnet} to train the regression head. 
Given two sentences $x_i,x_j$, the expression for the contrastive loss is $L_{x_i,x_j}^{con} = $max$(0,\vert d_{x_i,x_j}-\hat d_{x_i,x_j}\vert - \alpha)$, where $d_{x_i,x_j}$ and $\hat d_{x_i,x_j}$ respectively represent the difference of the labeled and predicted scores bewteen these two sentences. 
%The expressions for the contrastive loss is $L_{x_i,x_j}^{con} = $max$(0,\vert d_{x_i,x_j}-\hat d_{x_i,x_j}\vert - \alpha)$, given two samples $x_i,x_j$ from the same batch, $d_{x_i,x_j}$ represents the difference in the scores of these two utterances, $\hat d_{x_i,x_j}$ represents the difference in the predicted scores of these two utterances. 
$\alpha>0$ is a hyperparameter. 
The contrastive loss aims to penalize cases where the ranking between the predicted scores of $x_i$ and $x_j$ differs from their actual labels. 
The clipped loss is defined as $L^{clip}(y_i,\hat y_i) = \mathbb{I}(\vert y_i -\hat y_i \vert > \tau)(y_i - \hat y_i)^2 $, expecting to mitigate overfitting issues, where $y_i$ and $\hat y_i$ are the labeled and predicted scores of sentence $x_i$.  
$\mathbb{I}$ is indicator function and $\tau$ is a hyperparameter.
%The expressions for the clipped loss is $L^{clip}(y,\hat y) = \mathbb{I}(\vert y -\hat y \vert > \tau)(y - \hat y)^2 $.
The overall loss function is expressed as $L = \beta{L^{clip}} + \gamma{L^{con}}$, where $\beta$ and $\gamma$ are hyperparameters.

\vspace{-1.5mm}
\subsubsection{Classification head} 
\vspace{-1.5mm}
MOS prediction can also be viewed as a classification task, predicting the probability of the score falling into various score ranges. 
Thus, we also introduce a classification head. 
The input is first processed through multiple linear layers, then averaged over all frames, and finally passed through a softmax layer to output a sentence-level probability vector with a length of 5.
%, which consists of linear layers and a softmax layer, outputting an utterance-level probability vector with a length of 5. 
Assuming the vector is $[p_1,p_2,p_3,p_4,p_5]$, the classification score is the expectation $c = \sum_{i=1}^{5} i \times p_i$. 
%by multiplying each element in the vector by weights. 
When the rater ID is not the mean-listener, the label representing the sample is the score given by the individual rater (an integer $i$ from 1 to 5). 
In this case, the target is a one-hot vector with a length of 5 with the $i$-th element being 1 and all other elements being 0. 
When the ID is the mean-listener, the label is $\frac 1M\sum_{m=1}^M label^m$, where $label^m$ is the one-hot vector representation of the rating given by the $m$-th listener and $M$ is the number of listeners.  
We use the cross-entropy loss to train the classification head after finishing the training of the regression head. 
%Converting MOS prediction into a classification problem is challenging to train, so we train the classification head separately after training the regression head.

\vspace{-1.5mm}
\subsection{Aggregation layer}
\vspace{-1.5mm}
%We further introduce a refinement layer to integrate the regression score $r$ and classification score $c$ to a final score $s$. 
%After the training in the second stage, we obtain two output scores from the base MOS predictor. 
Previous work \cite{gyires2022improving} has indicated that the scores outputted by the classification head and regression head performed inconsistently in different score ranges. 
Therefore, simple addition and averaging are not ideal. 
Thus, we add a linear layer to aggregate score $r$ and $c$ to a final score $s$, enabling the layer to autonomously learn the relationship between these two scores. 
The training criterion is to minimize the mean square error (MSE) between the score $s$ outputted by the aggregation layer and the labeled one.
%it can combine the advantages of the classification head and regression head, reducing the overall mean squared error (MSE) to some extent.

\begin{table*}[t]
\setlength{\belowcaptionskip}{-0.4cm}
\renewcommand{\arraystretch}{0.8}
 \caption{Experimental results of the abaltion studies evaluated on the BVCC and BC2019 datasets.}
 \label{table:ablation study}
 \centering
 \begin{tabular}{l| c c c c| c c c c} 
 
  \hline
  \multirow{2}{*}{Model} &\multicolumn{4}{c|}{BVCC}&\multicolumn{4}{c}{BC2019}\\
  \cline{2-9}
   & S\_MSE $\downarrow$ & S\_LCC $\uparrow$ & S\_SRCC $\uparrow$ & S\_KTAU $\uparrow$ &
   S\_MSE $\downarrow$ & S\_LCC $\uparrow$ & S\_SRCC $\uparrow$ & S\_KTAU $\uparrow$ \\  
  \hline
  SAMOS & 0.097 & 0.944 & 0.942& 0.797& 0.179 & 0.968 & 0.976& 0.895\\
  - semantic module & 0.270 & 0.822 & 0.822& 0.636& 0.511 & 0.798 & 0.848& 0.662\\
  - acoustic module & 0.088 & 0.935 & 0.931& 0.783& 0.223 & 0.962 & 0.960& 0.846\\
  - ID embedding & 0.114 & 0.928 & 0.929& 0.780& 0.220 & 0.966 & 0.971& 0.871\\
  - weight branch & 0.096 & 0.943 & 0.937& 0.791& 0.263 & 0.962 & 0.983& 0.908 \\
  - regression head & 0.136 & 0.925 & 0.926& 0.773& 0.355 & 0.959 & 0.962& 0.790\\
  - classification head& 0.116 & 0.944 & 0.942& 0.796& 0.207 & 0.969 & 0.975& 0.889\\
  - aggregation layer & 0.102 & 0.944 & 0.942& 0.797& 0.250 & 0.963 & 0.975& 0.895\\
  \hline
 \end{tabular}
\end{table*}

\vspace{-1.5mm}
\section{Experiment Setup}
\vspace{-1.5mm}
\subsection{Dataset}
\vspace{-1.5mm}
In this paper, the experiments followed the same settings as the VoiceMOS Challenge 2022 \cite{huang2022voicemos}. 
The dataset includes BVCC dataset \cite{cooper2022generalization} from the main track and BC2019 dataset \cite{wu2019blizzard} from the out-of-domain (OOD) track. 
The BVCC dataset contains 7,106 English utterances, with the training/development/test sets split in a ratio of 70$\%$/15$\%$/15$\%$. 
The data comes from participating systems from past Blizzard Challenges (BCs), the Voice Conversion Challenges systems, and samples generated by ESPNet \cite{watanabe2018espnet}. 
Each sentence in BVCC has scores from 8 listeners, which we utilized along with the initial average score. 
The BC2019 dataset consists of Mandarin utterances from the BC 2019, with 136, 136, and 540 utterances forming the training set, development set, and test set, respectively.
%with 136 samples for training, 136 samples for development, and 540 samples for test, as well as 540 unlabeled training samples that we did not use during training. 
Each sentence in BC2019 is scored by 10 to 17 listeners. 
Since the raters in BVCC and BC2019 datasets are different, we defaulted to considering the listeners as mean ones, when finetuning on BC2019. 

\vspace{-1.5mm}
\subsection{Evaluation metrics}
\vspace{-1.5mm}
We used MSE, linear correlation coefficient (LCC), spearman rank correlation coefficient (SRCC), and kendall rank correlation coefficient (KTAU) to evaluate MOS prediction models. 
These metrics focus on different aspects. 
In challenges like BC or VCC, we are more interested in the ranking of participating systems, hence metrics like SRCC and KTAU are more important, while MSE is more reasonable when evaluating individual synthetic systems. 
In the VoiceMOS Challenge 2022 \cite{huang2022voicemos}, the organizers used system-level SRCC to determine rankings. 
%Considering the future applications of MOS automatic evaluation, MOS prediction models need to accurately assess the quality of multiple systems and evaluate different systems at an overall level. 
Considering that MOS automatic prediction models are mostly used to compare the quality differences of generated speech between systems, we used system-level (indicated by the prefix S\_) metrics as the standard to evaluate different MOS prediction models in the experiments.

\vspace{-1.5mm}
\subsection{Implementation}
\vspace{-1.5mm}
For the feature extractor, we used a pre-trained BiVocoder on the VCTK-0.92 corpus \cite{yamagishi2019cstr} and a 1-layer Conformer in the acoustic module. 
The pre-trained wav2vec2.0 from fairseq constituted the semantic module. 
The output feature dimensions of the semantic and acoustic modules were 64 and 768, respectively. 
The ID embedding dimension was set to 128. 
For the base MOS predictor, 3 BiLSTM layers with 128 nodes were used. 
In the regression head, both the frame-level score prediction and the weight prediction used 2 linear layers, and we set $\alpha = 0.1$, $\beta = 1$, $\gamma = 0.5$, and $\tau = 0.25$. 
In the classification head, 2 linear layers were adopted. 
% For the acoustic module, a subset of the VCTK-0.92 corpus \cite{yamagishi2019cstr} was used for pretraining BiVocoder. The number of layers of conformer is 1. For the ID embedding, the embedding dimension is 128. For the semantic module, we used the pre-trained wav2vec2.0 from fairseq%\footnote{https://github.com/facebookresearch/fairseq/tree/main/examples/wav2vec}%, as used in SSL-MOS
% . For the base MOS predictor, the number of LSTM layers is 3. For the loss function of the regression head, we set $\alpha = 0.1$, $\beta = 1$, $\gamma = 0.5$, $\tau = 0.25$. 
During training on BVCC, we trained for 1000 epochs, with a batch size of 8 and optimizer as stochastic gradient descent (SGD) with a learning rate of 0.0001. 
For the checkpoint saving strategy, we followed the same approach as UTMOS \cite{saeki2022utmos}, selecting the best system-level SRCC checkpoint calculated from the development set. 
If the system-level SRCC didn't decrease within 15 epochs, early stopping was applied. 
Since calculating metrics based on a single checkpoint may have some randomness, we choosed to average the parameters of the three best checkpoints to obtain a new model for testing. 
Due to the limited training data in the OOD track, we fine-tuned the SAMOS model trained on BVCC using BC2019 dataset. 

%\vspace{-1.5mm}
\subsection{Baselines}
%\vspace{-1.5mm}
We adopted the baseline SSL-MOS from the VoiceMOS Challenge 2022 and some excellent participating systems as baselines in the experiments. 
%SSL-MOS was the baseline of the VoiceMOS Challenge 2022, ranking 2nd in the OOD track in terms of two system-level metrics. 
Specifically, UTMOS used ensemble learning, integrating the neural network-based UTMOS strong with some machine learning methods. 
UTMOS ranked 1st, 2nd, 3rd, 3rd in the four system-level metrics in the main track, and 1st in all four system-level metrics in the OOD track. 
Currently, UTMOS strong without a phoneme encoder, is widely used for automatic MOS scoring for synthetic systems. 
Therefore, we separately considered UTMOS strong as a new baseline for comparison. 
DDOS added a classification head based on SSL-MOS, ranking 2nd in three out of four system-level metrics in the main track and 3rd in one remaining metric. 
However, DDOS performed poorly in the OOD track. 
Team T11 integrated multiple SSL models and ASR models, ranking 1st in three system-level metrics in the main track and 1st in two system-level metrics in the OOD track. 

%\vspace{-1mm}
\section{Results and Analysis}
%\vspace{-1mm}
\subsection{Comparision with baseline methods}

We first compare the proposed SAMOS with the baselines.
As shown in Table \ref{table:main experiment}, the experimental results on the BVCC dataset indicated that the proposed SAMOS significantly outperformed baseline models on three system-level metrics emphasizing correlation. 
Compared to SSL-MOS, it also reduced system-level MSE. 
On BC2019 dataset, the SAMOS ranked 2nd in two metrics emphasizing the correctness of system rankings, just behind UTMOS. 
However, UTMOS adopted an ensemble strategy, integrating over a hundred learner. 
This method requires extensive computational resources and is not conducive to practical application.
%Since UTMOS is the result of integrating over a hundred learner, and model ensembles generally consume time and resources, we believe that using a MOS prediction network with multiple model ensembles is not convenient for practical application. 
Under the condition that no ensemble strategy is used, compared to the single-neural-network-based UTMOS strong, SAMOS surpassed it on three metrics. 
Therefore, our proposed SAMOS also demonstrates comparable performance to the baselines on OOD data, confirming the robustness and stability of our proposed method.
%Therefore, we believe that after fine-tuning on a small amount of OOD Chinese data, SAMOS also has certain performance and efficiency advantages compared to existing models.

\vspace{-1mm}
\subsection{Ablation studies}
\vspace{-1mm}
We then conduct ablation experiments on SAMOS to explore the roles of each component. 
%We separately examined the impact of each module on performance. 
The results are shown in Table \ref{table:ablation study}. 
We first investigate the contribution of the semantic information, acoustic information, and listener ID informantion to the overall model performance. 
We can see that removing the semantic module resulted in the degradation of all the metrics on both datasets, indicating the importance of semantic representations from SSL model. 
When we removed the acoustic module, the results showed that the acoustic features extracted by BiVocoder was also indispensable. 
Removing the ID embedding, although causing a performance drop, still outperformed the original baseline SSL-MOS on the BVCC dataset. 
This indicates that acoustic features can effectively compensate for the performance loss caused by the absence of ID information.
% Removing the ID embedding led to results lower than SAMOS but still superior to the original baseline SSL-MOS on the English dataset, suggesting the feasibility of incorporating acoustic information into SSL-MOS when ID information is missing from the dataset. 
% Then, removing the weight branch performed lower than SAMOS on most metrics, but its metrics related to ranking on BC2019 were higher than SAMOS and comparable to UTMOS. 
The weight branch in the regression head was also proven effective, as hypothesized in Section \ref{subsec: Regression}.
%Subsequently, we removed the regression head and the classification head separately. 
Subsequently, we ablated the entire regression head, resulting in a decrease in all metrics to some extent. 
This indicates that treating MOS prediction as a classification task is challenging. 
% Removing the regression head resulted in a decrease in all metrics to some extent, indicating that treating MOS prediction simply as a classification problem is not feasible. 
Finally, Removing the classification head and aggregation layer both led to a decrease in MSE, indicating that the multi-task head framework had lower error compared to a single regression head. 
Using an aggregation layer instead of directly adding the scores from the two heads also improved prediction accuracy.%Removing the classification head led to a deterioration in the MSE metric, as shown in Figure \ref{fig:mse}, because the two heads have different abilities to predict MOS scores in different score ranges. The refinement layer combines the scores output by the two heads, ensuring that the predicted scores fall within the score ranges where the real scores appear most frequently, thereby reducing the overall MSE. ‘‘- Refinement layer" indicated that removing the refinement layer would increase the MSE.

\vspace{-1mm}
\section{Conclusions}
\vspace{-1mm}

% This paper proposes a MOS prediction model named SAMOS that simultaneously utilizes semantic and acoustic information. The model leverages comprehensive magnitude and phase features extracted by a BiVocoder based on existing models. Our experimental results demonstrate that SAMOS outperforms existing models on major English datasets and is comparable in performance to existing models after fine-tuning with a small amount of Chinese data. 

This paper presents a novel MOS prediction model called SAMOS which simultaneously utilizes semantic and acoustic information as input. 
To improve prediction accuracy, SAMOS employs parallel regression and classification heads, and finally outputs the final MOS score through an aggregation layer. 
Experimental results confirm that our proposed SAMOS significantly outperforms other baselines on system-level metrics, especially on the main track English dataset. 
Applying SAMOS to actual speech generation systems and guiding their training will be our future work.

\bibliographystyle{IEEEtran}

\bibliography{mybib}

% \begin{thebibliography}{9}
% \bibitem[1]{Davis80-COP}
%   S.\ B.\ Davis and P.\ Mermelstein,
%   ``Comparison of parametric representation for monosyllabic word recognition in continuously spoken sentences,''
%   \textit{IEEE Transactions on Acoustics, Speech and Signal Processing}, vol.~28, no.~4, pp.~357--366, 1980.
% \bibitem[2]{Rabiner89-ATO}
%   L.\ R.\ Rabiner,
%   ``A tutorial on hidden Markov models and selected applications in speech recognition,''
%   \textit{Proceedings of the IEEE}, vol.~77, no.~2, pp.~257-286, 1989.
% \bibitem[3]{Hastie09-TEO}
%   T.\ Hastie, R.\ Tibshirani, and J.\ Friedman,
%   \textit{The Elements of Statistical Learning -- Data Mining, Inference, and Prediction}.
%   New York: Springer, 2009.
% \bibitem[4]{YourName17-XXX}
%   F.\ Lastname1, F.\ Lastname2, and F.\ Lastname3,
%   ``Title of your INTERSPEECH 2022 publication,''
%   in \textit{Interspeech 2022 -- 23\textsuperscript{rd} Annual Conference of the International Speech Communication Association, September 18-22, Incheon, Korea, Proceedings, Proceedings}, 2022, pp.~100--104.
% \end{thebibliography}

\end{document}